\documentclass[aps,pre,twocolumn,groupedaddress,showpacs,amsfonts,10pt]{revtex4}
\usepackage{graphicx}
\usepackage{amsmath}
\usepackage{amssymb}

\usepackage{color,colordvi}
\definecolor{Red}{rgb}{0.9,0.0,0.1}

\begin{document}

%Title of paper
\title{Localization transition of stiff directed {lines} in random media}

\author{Horst-Holger Boltz}
\email[]{horst-holger.boltz@udo.edu}

\author{Jan Kierfeld}
\affiliation{Physics Department, TU Dortmund University, 
44221 Dortmund, Germany}

\date{\today}

\begin{abstract}
We investigate the localization of stiff directed {lines} 
 with bending energy by a short-range random potential. 
Using perturbative arguments, Flory arguments, and a replica 
calculation, we show that a stiff
 directed  {line} in 1+d  dimensions
undergoes a localization transition 
with increasing disorder  for $d>2/3$. 
We demonstrate that this transition
 is  accessible by numerical transfer matrix calculations
 in $1+1$ dimensions and analyze 
the properties of the disorder-dominated phase. 
{On the basis of the two-replica problem,} we propose a relation between 
 the localization of stiff directed {lines} in $1+d$ 
dimensions and of  directed {lines} under tension 
in $1+3d$  dimensions, which is strongly supported by identical free 
energy distributions. 
This shows  that  pair interactions in the replicated 
Hamiltonian  determine the nature of directed {line}
 localization transitions
with consequences for the critical behavior of the 
Kardar-Parisi-Zhang (KPZ) equation. 
Furthermore, we quantify how the persistence length 
of the stiff directed {line} is reduced by disorder. 
\end{abstract}


%02.50.-r Probability theory, stochastic processes, and statistics 
%05.40.-a Fluctuation phenomena, random processes, noise, and Brownian motion 
%64.70.-p 	Specific phase transitions
%64.60.Ht 	Dynamic critical phenomena
%61.41.+e 	Polymers, elastomers, and plastic

\maketitle

%%%%%%%%%%%%%%%%%%%
%\paragraph*{Introduction.}

Directed {lines (DLs)}
 in random media or the more general problem of 
random elastic manifolds are  one of the most important  
model systems in the statistical physics of disordered systems
\cite{HalpinHealy1995}. 
DLs in random media are related to  important 
non-equilibrium statistical physics problems 
such as stochastic growth, in particular 
the KPZ equation \cite{Kardar1986},
 Burgers turbulence, or the asymmetric 
simple exclusion model (ASEP) \cite{Krug1997}.
 Furthermore, there are many applications such as 
 kinetic roughening \cite{Krug1997},
pinning of flux lines \cite{Blatter1994, Nattermann2000}, 
random magnets, or wetting fronts \cite{HalpinHealy1995}. 

\begin{figure}
  \includegraphics[width=0.49\textwidth,height=0.2\textwidth]{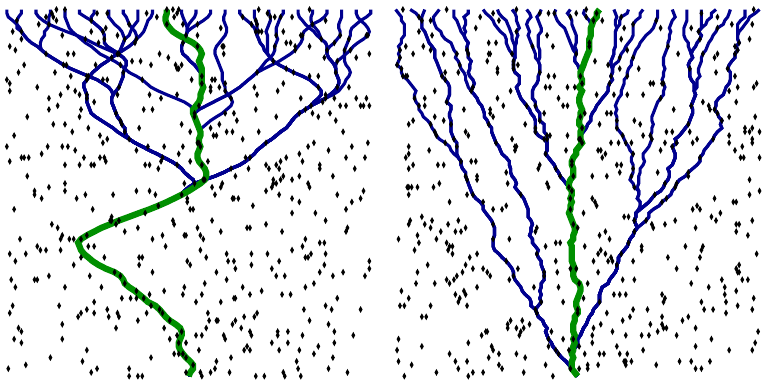}
 \caption{(color online) Paths with lowest energy for given ending states
   (top, globally optimal path thicker) for (a) the stiff  and (b) 
    the tense
   directed line (right). %The minimization of the energy corresponds to 
%$T=0$. 
The dots mark  favorable
   regions of the Gaussian random potential $V$
  (realizations of  the quenched
   disorder are not identical in a and b).} 
 \label{fig:t0}
\end{figure}

DLs  in
$D=1+d$-dimensional random media exhibit a
disorder-driven localization transition for dimensions
$d>2$, which has been 
studied numerically for dimensions up to  $d=3$ 
\cite{Derrida1990,Kim1991,Kim1991b,Monthus2006a}.
In the context of the KPZ equation, it  is a long-standing 
open question whether there exists an upper  critical dimension,
where the critical behavior at the localization transition is modified
(for a recent discussion, see, e.g., \cite{Katzav2002}).
We will address this question from a new perspective by 
revealing a relation between DLs in $1+3d$ dimensions
and {\em stiff} 
directed {lines (SDLs)}  in $1+d$ dimensions in a random 
medium, which we validate for $d=1$ by numerical simulations. 
We define SDLs as lines with 
 preferred orientation and  no overhangs with respect to this
direction, but a different elastic energy penalizing curvature instead of
stretching, cf.\  Fig.\ \ref{fig:t0}. We investigate their disorder-induced
localization transition and the  scaling properties of conformations 
 in the disordered phase.
 Because the proposed relation between DLs and SDLs is based on 
return probabilities of {\em  pairs} of replicas our results 
 suggest that the 
critical  properties of DLs in a random potential and, thus,
of the KPZ equation are determined by  the corresponding 
two replica problem.

%%%%%%%%%%%%%%%%
\paragraph*{Model of the stiff directed {line}.}

The configuration of a directed {line}, i.e.\ one without overhangs or loops
{and without inextensibility constraint}, 
can be parametrized by $(x,{\bf z}(x))$
$(0\leq x\leq L)$ with a $d$-dimensional displacement 
 ${\bf z}(x)$  normal to its preferred orientation. 
Each configuration of a SDL is associated with an energy 
\begin{align}
 {\cal H} &= 
  \int_0^L \mathrm{d}x\,\left[\frac{\kappa}{2} (\partial_x^2 {\bf z}(x))^2 
          + V(x, {\bf z}(x))\right], 
\label{eq:ham}
\end{align}
where the first term is the bending energy (to leading order in ${\bf z}$). 
The second term is the disorder energy with a 
Gaussian distributed 
quenched  random potential $V(x,{\bf z}(x))$ 
 with zero mean $\overline{V}=0$ and
short-range correlations
$\overline{V(x,{\bf z})V(x',{\bf z}')} = 
  g^2\delta^d({\bf z}-{\bf z}')\delta(x-x')$
 \footnote{$\overline{X}$ denotes the quenched disorder average over
  realizations of $V$, whereas ${\langle}X{\rangle}$ denotes  thermal
  averaging.}.

The SDL model \eqref{eq:ham} is an approximation
to the so-called worm-like chain or Kratky-Porod
model \cite{Harris1966,Kratky1949} ${\cal H}_{\text{WLC}} =
\int_0^L\mathrm{d}s\, \left[\frac{\kappa}{2} (\partial_s^2 {\bf r}(s))^2 
          + V({\bf r}(s))\right]$
which is a common model for semiflexible polymers, 
such as 
DNA or cytoskeletal filaments like F-actin. The chain is
parametrized in arc length, leading to a local inextensibility constraint
$\lvert \partial_s {\bf r}(s)\rvert=1$, which is hard to account for, both
numerically and analytically \cite{Kleinert2006,Dua2004}. 
{The approximation} \eqref{eq:ham} {only} applies to a weakly bent  
semiflexible polymer on length scales
{\em below}
 the so-called persistence length, which is \footnote{We use energy units
$k_B\equiv 1$.} $L_p =(D-1)\kappa/2T$ 
for thermally fluctuating semiflexible
polymers \cite{Kleinert2006}. 
Above the persistence length, a semiflexible polymer 
loses orientation correlations and starts to develop 
overhangs. Also in a quenched random potential the SDL model
describes semiflexible  polymers 
in heterogeneous media, {only}  as long as 
tangent fluctuations are small such that overhangs can be
neglected, which  is the case below a {\em disorder-induced} persistence 
length, which we will derive below.  

We consider the SDL model also in the
thermodynamic limit beyond this persistence length, 
because  we find evidence for   a relation to 
 the important problem of DLs in a random medium 
in {\em lower} dimensions. {This relation 
is based on replica pair interactions and 
shows that pair interactions also determine the nature of the
DL localization transition. Moreover, this relation can make
the DL transition in high dimensions  
computationally accessible. }
  We will now outline the idea behind 
this relation.

%%%%%%%%%%%%%%%%%%%%%%%
\paragraph*{Relation to directed lines.}

The  difference between  SDLs and   DLs
  \cite{HalpinHealy1995} is the {\em second} derivative in the
first bending energy term in eq.\
(\ref{eq:ham}) for SDLs, which differs from  the  tension
or stretching energy
$\sim \int dx (\partial_x {\bf z}(x))^2$ of DLs. 
This results in  different types of energetically favorable configurations
 (see Fig.\ \ref{fig:t0}): large perpendicular 
displacements ${\bf z}$ are not 
unfavorable as long as their ``direction'' does not change. 
Displacements ${\bf z}$ are characterized by the 
 {\em roughness exponent} $\zeta$, which is  defined by
$\overline{\langle z^2(L)\rangle}\sim L^{2\zeta}$.
The  thermal roughness is   $\zeta_{th,\tau}=1/2$ for  DLs and
$\zeta_{th,\kappa}=3/2$ for SDLs. 
Here and in the following we
use subscripts $\tau$ (tension) and $\kappa$ (bending stiffness) 
to distinguish between the two systems.

Although typical configurations are quite different,
a SDL subject to a short-ranged 
(around $z=0$) attractive potential
$V({\bf z})$ can be mapped onto a DL in high dimensions
$d'=3d$ \cite{Bundschuh2000,Kierfeld2005}. 
This equivalence is based on the probability that a
free line of length $L$ starting at
(${\bf z}(0)=0$) ``returns'' to the origin, i.e., ends at ${\bf z}(L)=0$.
This return probability is characterized by a {\em return  exponent} $\chi$: 
$\text{Prob}({\bf z}(L)=0)\sim L^{-\chi}$.
For DLs, which are  essentially random walks in $d$ transverse
dimensions, the return exponent is $\chi_{\tau}=d/2$ \cite{Fisher1984}, 
whereas it is $\chi_{\kappa}=3d/2$ for a SDL (after 
integrating over all orientations of the end) \cite{Gompper1989}; 
they are related to the roughness exponents by $\chi= \zeta d$
\cite{Kierfeld2005}.
The return exponent 
governs the binding to a short-range attractive potential 
or, equivalently, the binding of two lines interacting by such a potential
\cite{Fisher1984, Bundschuh2000, Kierfeld2005}.
The relation $\chi_{\tau}(3d)=\chi_{\kappa}(d)$ 
implies that the binding transition 
of two DLs in $3d$ dimensions maps onto 
to the binding transition of two SDLs in $d$ 
dimensions. 

In the replica formulation {of line localization problems such as 
 (\ref{eq:ham})},
the random potential gives rise to a short-range 
attractive pair interaction (see below). 
Furthermore, the critical temperature
$T_{c,\tau}$ for a DL in a random potential is
believed to be identical to the critical
temperature $T_{2,\tau}$ for a system with two replicas
\cite{Monthus2006a, Monthus2007}. One aim of this
work is to generalize this conjecture by demonstrating 
that the $d\rightarrow 3d$ analogy between DLs and SDLs in 
a random potential holds for the {\em entire}
free energy distribution. Because this analogy is rooted in 
the binding transition of replica pairs, we can conclude that
 critical properties of the localization transition are 
determined by  pair interactions in the replicated 
Hamiltonian, which  has  been previously 
suggested in Refs.\ \cite{Bundschuh1996,Mukherji1996}. 
Moreover, it has been proposed  that 
pair interactions can be used to formulate an 
 order parameter of the disorder-driven localization 
transition of DLs in terms of the overlap 
$q\equiv \lim_{L\rightarrow\infty}\overline{\frac{1}{L}\int_0^L\mathrm{d}x
\delta(z_1(x)-z_2(x))}$ \cite{Mezard1990,Mukherji1994,Comets2003}, i.e.,
 the average number of sites per length, that two lines in the same
realization of the disorder have in common.
Localization by disorder gives rise to a 
 finite value of the pair overlap $q$.

%%%%%%%%%%%%%%%%%%%%%%%%%%%%%%%
\paragraph*{Scaling analysis.}

\begin{figure}
\begin{center}
 \includegraphics{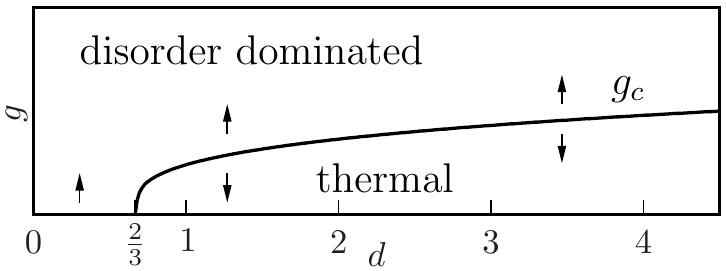}
 \caption{Sketch of the phase diagram as implied by Flory-type arguments.
The arrows indicate the flow of the disorder ``strength'' $g$ under
renormalization.}\label{fig:renorm}
\end{center}
\end{figure}

We start with a scaling analysis of SDL behavior in disorder 
 by a Flory-type argument. 
For displacements $\sim z$ the bending energy  in eq.\
(\ref{eq:ham}) scales as $E_b \sim \kappa z^2 L^{-3}$, which also leads to 
$\langle z^2 \rangle \sim  L^3/L_p$ and the thermal roughness exponent
$\zeta_{th,\kappa}=3/2$.
The disorder energy in eq.\  (\ref{eq:ham})
 scales as $E_d \sim g\sqrt{L z^{-d}}$. 
Using the  unperturbed  thermal roughness in the disorder energy 
we get  $E_d \sim L^{(2-3d)/4}$, 
from which we conclude that the disorder
is relevant below a critical dimension
 $d<d_{c,\kappa}=2/3$. 
For  $d>d_{c,\kappa}=2/3$ and, thus, 
in all physically accessible integer 
dimensions, the SDL should exhibit a transition 
from  a thermal phase for low $g$
 to a disorder dominated phase above a  critical value $g_c$
of the disorder (see Fig.\ \ref{fig:renorm}). 
In the disorder dominated phase, the SDL
 becomes localized 
%into a configuration favored by the random potential
and assumes a roughened configuration, see Fig.\ \ref{fig:t0}.

Balancing the Flory estimates, $E_b \sim E_d$,  gives the roughness 
  $z\sim L^{\zeta_{Fl}}$, if disorder is relevant. This leads to 
 $\zeta_{Fl,\kappa}=7/(4+d)$, which 
is applicable below the critical dimension $d<d_{c,\kappa}$,
where $\zeta_{Fl,\kappa}>\zeta_{th,\kappa}$.
Above the critical dimension, the Flory result would give  
 $\zeta_{Fl,\kappa}<\zeta_{th,\kappa}$, which  contradicts 
a roughening of  the SDL as it 
 adjusts to the random potential. 
Furthermore, the exponent $\omega$ related to the energy
fluctuations via $\Delta F \sim L^{\omega}$
would be negative, since the scaling of the energy implies a general
scaling relation $\omega=2\zeta_{\kappa}-3$ (note that we do not subscript
$\omega$ as we believe $\omega_{\tau}=\omega_{\kappa}$, see below).
An exponent $\omega<0$  contradicts 
the existence of  large disorder-induced free energy fluctuations 
 in the low-temperature phase \cite{Fisher1991,Monthus2006b},
for which there is also strong numerical evidence  
\cite{Kim1991,Kim1991b,Monthus2006a}.
We conclude that this kind of argument is
not applicable above the critical dimension.
The same problems  occur in Flory arguments 
for DLs for $d>d_{c,\tau}=2$ as well as in, for example, functional
renormalization group analysis \cite{LeDoussal2005} for $d\gtrapprox 2.5$.
We conclude that, {in contrast to DLs,}
 it is sufficient to study 1+1 dimensional 
SDLs in numerical transfer matrix calculations in order to 
explore the properties of their 
localization transition  such that we 
focus on this case in the following.

%%%%%%%%%%%%%%%%%%%%%%%%%%%%%%%
\paragraph*{Variation in replica space.}

To go beyond  scaling arguments we use the replica
technique \cite{Dotsenko2001} following the treatment of directed
manifolds \cite{Mezard1991}, the results of which
we summarize briefly in the following.
We write the  averaged and  replicated partition function as
$\overline{Z^n}=\prod_\alpha(\int\mathcal{D}z_\alpha)\exp{(-\beta
  {\cal H}_{\text{rep}})}$ with the replica Hamiltonian in Fourier
space $  {\cal H}_{\text{rep}}=\frac{1}{2L}\!\sum_{\!\alpha=1\!}^n\!
     \sum_k (\kappa k^4{+} \mu) z_\alpha^2{-} \frac{\beta g^2}{2}\!
\sum_{\!\alpha,\beta=1\!}^n \int_0^L\!\!
  \mathrm{d}x\,f_\lambda(z_{\alpha\beta}^2)$
where  $z_{\alpha\beta}\equiv z_{\alpha}-z_\beta$ and 
with an attractive potential $f_\lambda(z)$ of range $\lambda$. 
As mentioned before, ${\cal H}_{\text{rep}}$ is  related to a pair binding
problem: in the  limit $\lambda \approx 0$ the second
term becomes $- \frac{\beta g^2}{2} \sum_{\alpha,\beta} \int_0^L
\mathrm{d}x\delta(z_\alpha-z_\beta)$. 

Using variation in replica space we find  one-step replica symmetry breaking
for $d>d_{c,\kappa}$ and 
 can show that there is no localized 
solution unless the potential strength $g$
and correlation length $\lambda$ are above finite values. 
%Although variation in
%replica space is known to fail in reproducing the exact solution
%\cite{Kardar1987} for the problem of the DL  in disorder, 
We interpret this as
an indication for the existence of a critical disorder strength or a critical
temperature for $d>d_{c,\kappa}$.

%%%%%%%%%%%%%%%%%%%%%%%%%%%%
\paragraph*{Numerical results.}

\begin{figure*}[t]
\begin{center}
 \includegraphics[width=0.95\textwidth]{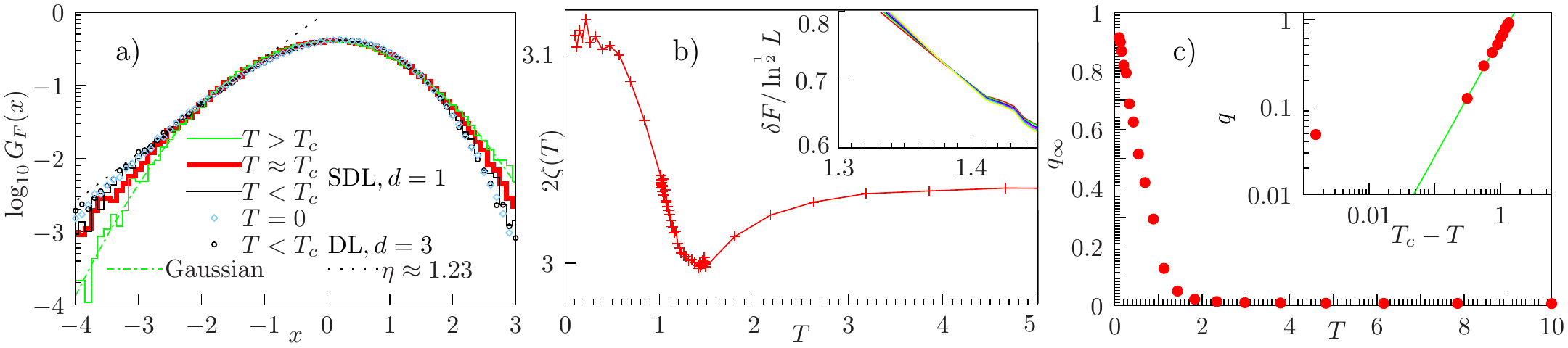}
 \caption{(color online) a): Rescaled $G_F(X)=P((F-\overline{F})/\Delta F)$
free energy distribution for a stiff directed line (SDL) in $1+1$
dimensions. 
  We show distributions for $T=0$ (light (blue) squares, ground state energy) as
well as
for three finite  temperatures $T<T_c$ (black thin solid line), $T\approx T_c$
(red, thicker solid line) and
   $T>T_c$ (green, light thin solid line). Results for a directed line (DL) 
  in 1+3 dimensions are shown (dark circles). b):
   Local (see text) roughness exponent $2\zeta_{\kappa}$ as a 
  function of $T$. Deviations
  from the analytical value $2\zeta_\kappa=3$ at high temperatures indicate
   numerical problems, nonetheless there is a clear ``dip'' at
   $T\approx 1.4$, which we identify as the critical temperature. For
   low temperatures, we find values consistent with
   $\omega\approx 0.11$.
    Inset: Reduced free energy 
    $\delta F=\overline{F}-\overline{F}_\text{ann}$
    rescaled by $\ln^{1/2}{L}$  for lengths
   $L=50,60,\ldots,100$ as a function of  $T$. There is
   a pseudocrossing around
   $T\approx1.38$. c): The overlap order parameter  $q$, as a
  function of $T$. We estimated $q$ from
 finite lengths using a fit $q_T(L)=a(T)/L+q_\infty(T)$.
 %, which is possibly not accurate for $T=T_c$ \cite{Mukherji1994}. 
Inset: Double-logarithmic plot
of the overlap $q$ versus  $T_c-T$ (with $T_c=1.44$), the
solid line is given by $q\sim (T_c-T)^{-\beta'}$ with $\beta'\approx-
1.36$.}
\label{fig:zeta}\label{fig:deltaf}\label{fig:g_f} 
\end{center}
\end{figure*}

More progress is possible by 
 extensive numerical studies using the 
 transfer matrix method \cite{HalpinHealy1995,Wang2000} both 
for $T=0$ (see Fig.\ \ref{fig:t0}) and for  $T>0$. The transfer matrix
element for a segment of a SDL with length $\Delta L =
1$ starting at $z$ with orientation $\mathrm{d}z/\mathrm{d}x=v$ and ending
at $z'$ with $v'$ follows from a discretization of
(\ref{eq:ham}) as $\Delta E=\kappa/2(v-v')^2+V(z')$, where we have
chosen $z'=z+v'$ to simplify the transfer matrix. For the sake of
simplicity we choose $g=1$ and vary the temperature.

In order to test the relation between DLs and SDLs, 
we compare free energy fluctuations 
for DLs in 1+3 and SDLs in 1+1 dimensions. 
Previous direct numerical studies of  DLs in 1+3 dimensions
found an  exponent $\omega\approx 0.18$
\cite{Kim1991b,Monthus2006a} in  the low temperature phase, 
the most precise value from kinetic roughening studies is
 $\omega\approx 0.186$ \cite{Marinari2000}.
We compare these values with own numerical transfer matrix calculations
for SDLs in 1+1 dimensions. 
We determine the exponent $\omega$ directly by fitting 
$\Delta F=({\overline{F^2}-\overline{F}{}^2})^{1/2}\propto L^\omega$, which
gives values 
$\omega\approx 0.16$ for DLs in 1+3 dimensions, 
or by studying the distribution of the free
energy shown in Fig.\ \ref{fig:g_f}a, which is obtained  by computing the
free energy for every sample and  rescaling to zero mean and
unit variance,
$G_F(x)=\text{Prob}((F-\overline{F})/\Delta F=x)$.
The asymptotic behavior of the negative tail of the rescaled free
energy distribution for low temperatures, which is of the form
$\ln{G_F(x)}\sim -\lvert x\rvert^\eta$ ($x<0$,$\lvert x\vert\gg1$),
allows us to  determine  the energy fluctuation exponent $\omega$
via the Zhang argument \cite{HalpinHealy1995} giving
$\eta=1/(1-\omega)$. We get $\eta\approx1.23$ and therefore
$\omega\approx 0.19$. These values agree  with the literature 
values for DLs in 1+3 dimensions. 

For direct 
comparison of the entire free energy
distributions of a SDL in 1+1  and a 
  DL  in 1+3 dimensions  
 we simulated {\em both} systems and find  that
the rescaled free energy
distributions in the low temperature phases 
have to be considered {\em identical} within 
numerical accuracy, see Fig.\ \ref{fig:g_f}a.
This strongly
supports the relation between  DLs in $1+3d$ 
dimensions and  SDLs in $1+d$ dimensions.

In a third approach, we can calculate  $\zeta_{\kappa}$  and 
$\omega = 2\zeta_{\kappa}-3$ by measuring a 
``local'' version of the roughness exponent \cite{Schwartz2012}
$ 2\zeta(L)=\log_5 (z^2(L)/z^2(L/5))$. The data shown in Fig.\ \ref{fig:zeta}b
shows two distinct regimes for high and low temperatures and a significant
``dip'' around $T=1.4$.  For
   low temperatures, values $2\zeta_{\kappa}\approx 3.11$ are consistent with
   $\omega\approx 0.11$.
As for  DLs \cite{Doty1992}, 
it can be argued that $\omega$ should vanish at
the transition, resulting in fluctuations of the free energy that
scale logarithmically with $L$,  $\Delta F\sim \ln^{1/2}{L}$ 
\cite{Monthus2006a}, resulting in a 
roughness exponent $\zeta_{\kappa} = 3/2$. 
This seems to hold, even though the numerical
value for high temperatures is slightly above $\zeta_{\kappa}=3/2$,
 which is strong evidence for  a phase transition at
$T_c\approx 1.4$.  We support this by studying the
difference of quenched and annealed free energies
$\delta F=\overline{F}-\overline{F}_\text{ann}$,
which should as well scale as $\delta F\sim\ln^{1/2}{L}$
at the
transition. We determined $F_{\text{ann}}$ by
simulating a
system without disorder and adding the contribution of the annealed
potential, $\overline{F}_{\text{ann}} = F_{g=0}-Lg^2\beta/2$, see Fig.\ 
\ref{fig:deltaf}b.

Finally, we identify an order parameter of the localization
transition. 
For DLs, the disorder-averaged overlap 
$q= \lim_{L\rightarrow\infty}\overline{\frac{1}{L}\int_0^L\mathrm{d}x
\delta(z_1(x)-z_2(x))}$ of two replicas  has been proposed as order parameter
\cite{Mezard1990,Mukherji1994}. Up to now, it has been numerically impossible 
 to verify
this order parameter for DLs in $d>2$ dimensions where a localization 
transition exists because the relevant $2d$-dimensional 
two replica phase space is too large. 
For SDLs, on the other hand, 
   the transition is numerically accessible already in 
1+1 dimensions and we show that the overlap $q$ 
is indeed a valid order parameter using an adaptation
of the transfer matrix technique from Ref.\ \cite{Mezard1990}, see
Fig. \ref{fig:g_f}c. This involves simulating {\em two} interacting SDLs,
therefore we can only use lengths up to $L=30$ and $10^3$ samples. 
For DLs, it has been
found  that the overlap at criticality decays as
$q\sim L^\Sigma$ with $\Sigma=-2\zeta=-(1+\omega)$ in $d=3$ 
\cite{Mukherji1994}. This has been
extended to
finite temperatures yielding $q\sim \lvert T-T_c\rvert^{-\nu\Sigma}$. Indeed, 
we find a
qualitatively similar behavior $q\sim \lvert T-T_c\rvert^{-\beta'}$ with 
 an exponent $\beta'\approx 1.3-1.4$.   Our best estimate for $\Sigma$ 
 is $\Sigma\approx -0.75$. For the  correlation length exponent $\nu$
we find values  $\nu\approx 2$ compatible with 
 the corresponding problem of DLs\cite{Monthus2006a,Monthus2007,Derrida1990};
such that 
our present results   deviate  
from $\beta'=\nu \Sigma$. {Because of  small 
simulation lengths  $L$ 
we do not conclude this deviation to be a definite statement against the
renormalization group results presented in Ref.\ \cite{Mukherji1994}.}
Nevertheless, the connection between DLs and SDLs provides the first system to
test
the proposed order parameter in a localization transition numerically 
and to determine the {otherwise inaccessible} exponents  $\beta'$ or
$\Sigma$.

%%%%%%%%%%%%%%%%%%%%%%%%%%%
\paragraph*{Disorder-induced persistence length.}

\begin{figure}[t]
\begin{center}
 \includegraphics[width=0.4\textwidth,height=0.19\textwidth]{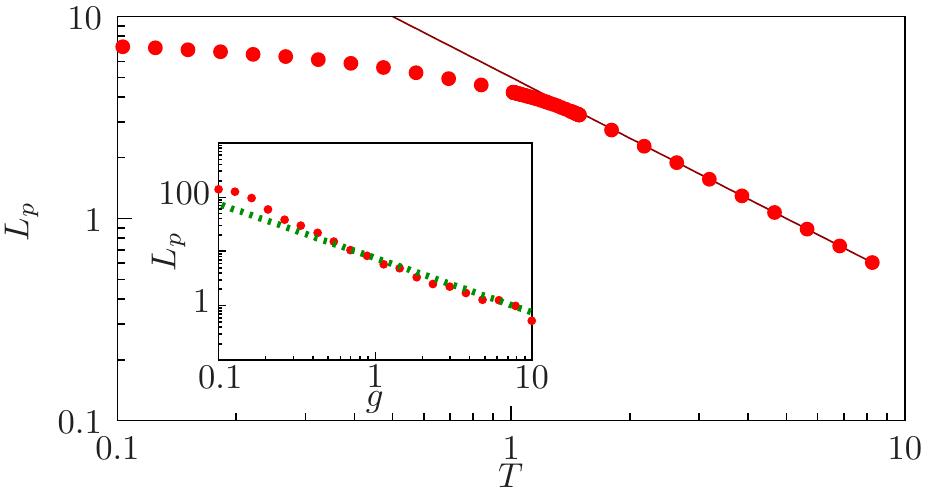}
 \caption{Disorder-reduced persistence length $L_p$ of the
SDL for $g=1$. 
$L_p$ matches its thermal
value (solid line) for $T>T_c$ and is reduced and approximately constant for
$T<T_c$. Inset:  $L_p$ at $T=0$ versus the potential strength
$g$. The Flory-result $L_p\sim g^{-1}$ (dotted line) 
matches the data.}
\label{fig:lp}
\end{center}
\end{figure}

As stated before, the model of a SDL in a random potential describes a 
worm-like chain in a heterogeneous environment on {small
 length scales $L<L_p$} such that overhangs can be neglected, which 
gives the defining 
criterion  $\overline{\langle v^2\rangle}(L=L_{p})=1$ for the 
persistence length.
 The line roughens in the low
temperature phase, which 
gives rise to  a persistence length {\em decrease} as compared to the 
thermal persistence length $L_p\sim \kappa/k_BT$. 
At low temperatures, the Flory-result $z\sim(g/\kappa)^{2/(4+d)}  L^{7/(4+d)}$
leads to $L_p \sim (\kappa/g)^{2/(3-d)}$.
% z\sim (g/\kappa)^{2/(4+d)}  L^{7/(4+d)} 
% v\sim (g/\kappa)^{2/(4+d)}  L^{(3-d)/(4+d)} 
For $d=1$, we determine 
$L_{p}$ numerically via the above defining criterion,
 and the data presented in Fig.\ \ref{fig:lp} does indeed show an
 only  weakly temperature-dependent disorder-reduced 
 persistence length for $T<T_c$. The $T=0$ results match 
the Flory-result $L_p \sim \kappa/g$.

%%%%%%%%%%%%%%%%%%%%%%%%%%%
\paragraph*{Conclusion.}

We studied {stiff directed lines (SDLs)}  in $1+d$ dimensions subject to 
 quenched  short-range random potential analytically and numerically. 
Using Flory-type scaling 
 arguments and a replica calculation we show that,
in dimensions $d>2/3$,  a localization transition exists 
from a high temperature
phase, where the system is essentially annealed, to a
disorder-dominated low temperature phase. 
 The low temperature  phase is characterized
by large  free energy fluctuations with  an exponent 
$\omega>0$, which cannot be 
calculated by scaling arguments or replica calculations. 
By extensive numerical transfer matrix 
calculations in 1+1 dimensions we find  a
value $\omega\approx 0.18$, which 
is close to the  established value $\omega \approx 0.186$ 
 for  directed lines (DLs) under
tension in 1+3 dimensions. Moreover, 
 the rescaled free energy distributions  are identical. 
Both points suggest that the nature of
the low-temperature phase is very similar, if not identical. 

This strongly supports a relation between DLs in $1+3d$
and SDLs in $1+d$ dimensions, which is based on identical  return 
exponents $\chi$ for two replicas to meet. The validity of 
a relation based on properties of a single replica pair suggests that 
the critical properties of  DLs in a short-range 
random potential are governed by replica pair interactions.
{
The mapping can make 
 DL transitions in high dimensions  computationally accessible,
 which we demonstrated in showing that the two-replica overlap 
provides a valid order parameter across the localization transition 
of SDLs in 1+1 dimensions. 
Furthermore, the importance of pair interactions suggests}
 that the critical temperature for DLs in random potentials 
is  indeed 
identical to the temperature below which the
ratio of the second moment of the partition function and the square of its
first moment diverges.
The binding transition of 
DL pairs becomes discontinuous for $d>4$  
and, analogously, the binding of  SDL pairs
for $d>4/3$  \cite{Kierfeld2003,Kierfeld2005}. 
Because DLs in random potentials are 
equivalent to  the KPZ equation \cite{Kardar1986}, the validated relation to
the SDL suggests that the roughening 
transition of the KPZ problem could acquire similar  
discontinuous features for $d>4$ dimensions. 
Finally, we calculated  the  reduction of the 
persistence length of a stiff directed by disorder. 

%%%%%%%%%%%%%%%%%%%%
We acknowledge financial support by  the Deutsche Forschungsgemeinschaft
(KI 662/2-1).

%%%%%%%%%%%%%%%%%%%%%%%%%%%%%%%%%%%%%%%%%%%%%%%%%%%%%%%%%%%%%%%%%%


\begin{thebibliography}{99}

\bibitem{HalpinHealy1995}%1
T. Halpin-Healy and  Y.C. Zhang, 
Phys. Rep. {\bf 254}, 215 (1995). 

\bibitem{Kardar1986}%2
M. Kardar,  G. Parisi, and  Y.C. Zhang, 
Phys. Rev. Lett. {\bf 56}, 889 (1986). 

\bibitem{Krug1997}%3
J. Krug, Adv. Phys. {\bf 46}, 139 (1997). 

\bibitem{Blatter1994}%4
G. Blatter, M.V. Feigelman, V.B. Geshkenbein, 
A.I. Larkin, and  V.M. Vinokur, Rev. Mod. Phys.  {\bf 66},  
1125 (1994). 

\bibitem{Nattermann2000}%5
T. Nattermann and  S. Scheidl, Adv.  Phys. {\bf 49}, 607 (2000). 

\bibitem{Derrida1990}%6
B. Derrida and  O. Golinelli, Phys. Rev. A {\bf 41}, 4160 (1990).

\bibitem{Kim1991}%7
J.M. Kim, A.J. Bray, and 
M.A. Moore, Phys. Rev. A {\bf 44}, R4782 (1991).

\bibitem{Kim1991b}%8
J.M. Kim, M.A. Moore, and A.J. Bray, Phys. Rev. A {\bf 44}, 2345 (1991).

\bibitem{Monthus2006a}%9
C. Monthus and  T. Garel, Eur. Phys. J. B {\bf 53}, 39 (2006).

\bibitem{Katzav2002}%10
E. Katzav and M. Schwartz, Physica A {\bf 309}, 69 (2002). 

\bibitem{Harris1966}%11
R.A. Harris and J.E. Hearst, J. Chem. Phys. {\bf 44}(7), 2595 (1966).

\bibitem{Kratky1949}%12
O. Kratky and G. Porod, Recl. Trav. Chim. Pays-Bas {\bf 68}(12), 1106 (1949).

\bibitem{Dua2004}%13
A. Dua, T. Vilgis, J. Chem. Phys. {\bf 121}, 5505 (2004).

\bibitem{Kleinert2006}%14
H. Kleinert, {\it Path Integrals in Quantum Mechanics, Statistics, Polymer
Physics, and Financial Markets} (World Scientific, Singapur, 2006). 

\bibitem{Bundschuh2000}%15
R. Bundschuh, M. L{\"a}ssig, and R. Lipowsky, 
Eur. Phys. J. E {\bf 3}, 295 (2000).

\bibitem{Kierfeld2005}%16
J. Kierfeld and  R. Lipowsky, 
J. Phys. A {\bf 38}, L155 (2005).

\bibitem{Fisher1984}%17
M.E. Fisher, J. Stat. Phys. {\bf 34}, 667 (1984).

\bibitem{Gompper1989}%18
G. Gompper and  T.W. Burkhardt, Phys. Rev. A {\bf 40}, 6124 (1989).

\bibitem{Monthus2007}%19
C. Monthus and  T. Garel, Phys. Rev. E {\bf 75}, 051122 (2007).

\bibitem{Bundschuh1996}%20
R. Bundschuh and M. L{\"a}ssig, Phys. Rev. E {\bf 54}, 304 (1996). 

\bibitem{Mukherji1996}%21
S. Mukherji and S.M. Bhattacharjee, Phys. Rev. B {\bf 53}, {R6002} (1996).

\bibitem{Mukherji1994}%22
S. Mukherji, Phys. Rev. E {\bf 50}, R2407 (1994).

\bibitem{Mezard1990}%23
M. Mezard, J. Phys. Frace {\bf 51}, 1831 (1990).

\bibitem{Comets2003}%24
F. Comets, T. Shiga and N. Yoshida, Bernoulli {\bf 9}(4), 705 (2003).

\bibitem{Fisher1991}%25
D.S. Fisher and D.A. Huse, Phys. Rev. B {\bf 43}, 10728 (1991).

\bibitem{Monthus2006b}%26
C. Monthus and  T. Garel, Phys. Rev. E {\bf 74}, 011101 (2006).

\bibitem{LeDoussal2005}%27
P. Le Doussal and K. J. Wiese, Phys. Rev. E {\bf 72}, 035101(R) (2005).

\bibitem{Dotsenko2001}%28
V. Dotsenko, {\it Introduction to the Replica Theory of Disordered Statistical
  Systems} (Cambridge University Press, Cambridge, 2001). 

\bibitem{Mezard1991}%29
M. Mezard and  G. Parisi, 
J. Physique I {\bf 1}, 809 (1991). 

\bibitem{Wang2000}%30
X. Wang, S. Havlin, and M. Schwartz, 
J. Phys. Chem. B {\bf 104}, 3875 (2000).

\bibitem{Marinari2000}%31
E. Marinari, A. Pagnani, and G. Parisi, J. Phys. A: Math. Gen. {\bf 33}, 8181
 (2000). The nomenclature differs: our $\omega$ is not the $\omega$ listed in
this paper, but $\omega=\frac{\chi}{2-\chi}$.

%\bibitem{Ala1993}
%T. Ala-Nissila, T. Hjelt, J.M. Kosterlitz, and 
%O. Ven{\"a}l{\"a}inen, J. Stat. Phys. {\bf 72}, 207 (1993). 

\bibitem{Schwartz2012}%32
M. Schwartz and E. Perlsman, Phys. Rev. E {\bf 85}, 050103(R) (2012).

\bibitem{Doty1992}%33
C.A. Doty and J.M. Kosterlitz, Phys. Rev. Lett. {\bf 69}(13), 1979 (1992).

\bibitem{Kierfeld2003}%34
J. Kierfeld and R. Lipowsky, Europhys. Lett. {\bf 62}, 285 (2003). 


%\bibitem{Parisi1980}
%G. Parisi, J. Phys. A {\bf 13}, 1887 (1980).

%\bibitem{Kardar1987}
%M. Kardar, Nucl. Phys. B {\bf 290}, 582 (1987). 


\end{thebibliography}
\end{document}